\documentclass[runningheads]{llncs}
\usepackage{graphicx}
\usepackage{multirow}
\usepackage{caption}
\usepackage{subcaption}
\begin{document}
\title{Brain Tumor Segmentation and Survival Prediction}
\titlerunning{Brain Tumor Segmentation and Survival Prediction}
%
\author{Rupal R. Agravat\inst{1}\orcidID{0000-0003-1995-4149} \and
Mehul S. Raval\inst{2}\orcidID{0000-0002-3895-1448}}
\authorrunning{Agravat R. and Raval M.}
%
\institute{Ahmedabad University, Ahmedabad, Gujarat, India \\
\email{rupal.agravat@iet.ahduni.edu.in}\\ \and
Pandit Deendayal Petroleum University, Gandhinagar, Gujarat, India\\
\email{mehul.raval@sot.pdpu.ac.in}}
\maketitle              
\begin{abstract}
The paper demonstrates the use of the fully convolutional neural network for glioma segmentation on the BraTS 2019 dataset. Three-layers deep encoder-decoder architecture is used along with dense connection at encoder part to propagate the information from coarse layer to deep layers. This architecture is used to train three tumor subcomponents separately. Subcomponent training weights are initialized with whole tumor weights to get the localization of the tumor within the brain. At the end, three segmentation results were merged to get the entire tumor segmentation. Dice Similarity of training dataset with focal loss implementation for whole tumor, tumor core and enhancing tumor is 0.92, 0.90 and 0.79 respectively. Radiomic features along with segmentation results and age are used to predict the overall survival of patients using random forest regressor to classify survival of patients in long, medium and short survival classes. 55.4\% of classification accuracy is reported for training dataset with the scans whose resection status is gross-total resection. 

\end{abstract}

\section{Introduction}

Early-stage brain tumor diagnosis can lead to proper treatment planning which improves patient survival chances. Out of all type of brain tumors, Glioma is one of the most life-threatening brain tumors. It occurs in glial cells of the brain. Depending on its severity and aggressiveness, it is divided into four grades ranging from grade I to grade IV(Grade I, II are Low-Grade Glioma(LGG) and grade III and IV are High-Grade Glioma(HGG)). A Brain tumor can further be divided into constituent components like - Necrosis, Enhancing tumor, Non-enhancing tumor and Edema. Tumor core consists of necrosis, enhancing tumor, non-enhancing tumor. In most cases, LGG doesn't contain enhancing tumor, whereas HGG contains necrosis, enhancing and non-enhancing subcomponent. Edema occurs from infiltrating tumor cells, as well as a biological response to the angiogenic and vascular permeability factors released by the spatially adjacent tumor cells\cite{akbari2014pattern}.

Quantification of tumor subcomponent plays an important role in whole tumor study and appropriate treatment planning. Non-invasive Medical Resonance Imaging(MRI) is the most advisable imaging technique as it captures the functioning of soft tissue properly compared to other imaging techniques. MR images are prone to inhomogeneity introduced by the surrounding magnetic field which introduces the artifacts in the captured image. In addition, the appearance of various brain tissue is different in various modalities. Such issues increase the time in the study of the image. And the human interpretation of the image is non-reproducible as well as dependent on the expertise. This requires computer-aided MR image interpretation to locate the tumor.  

Authores in \cite{agravat2018deep} classified brain tumor segmentation in basic, generative and discriminative techniques. Nowadays, Deep Neural Networks(DNN) has gained more attention for the segmentation of biological images. In which, Fully Convolution Neural Networks(FCNN), like U-Net\cite{ronneberger2015u}, V-Net\cite{milletari2016v}, SegNet\cite{badrinarayanan2017segnet}, ResNet\cite{he2016deep}, DenseNet\cite{iandola2014densenet} give state-of-the-art results for semantic segmentation. Out of all these methods, U-net is widely accepted end-to-end segmentation architecture for brain tumors. In\cite{kao2018brain} authors used ensemble of various DNN architectures and supplied and utilized brain parcellation atlas for brain tumor segmentation. Connectomics data, parcellation information and tumor mask were used to generare fetures for survival prediction. Authors of \cite{baid2018deep} supplied 3D patches to 3D U-net for tumor segmentation and used radiomics features for survival prediction. In\cite{chen2018drinet}, dense module, residual module and inception modules were used for biomedical image segmentation. In all the mentioned approaches, encoder-decoder architecture is used. 

According to \cite{agravat2019os}, inductive transfer learning\cite{pan2009survey} improves the network performance. In this paper we have implemented U-net\cite{dong2017automatic},\cite{ronneberger2015u} with reduced network depth, replaced convolution module at encoder path with dense module and used inductive transfer learning for initializing subregion network training. 

The remaining paper is organized as follows: section two of the paper focuses on the BraTS 2019 dataset, section three demonstrates the proposed method, section four provides implementation details and results. At last, the conclusion followed by future work is given. 

\section{Dataset}

The dataset \cite{bakas2017advancing},\cite{bakas2018identifying},\cite{menze2014multimodal} contains 259 HGG and 76 LGG pre-operative scans. To generate the ground truth, all the images have been segmented manually, by one to four raters, following the same annotation protocol, and their annotations were approved by experienced neuro-radiologists\cite{bakas2017segmentation},\cite{bakas2017segmentation1}. Annotations comprise the enhancing tumor (ET — label 4), the peritumoral edema (ED — label 2), and the necrotic and non-enhancing tumor core (NCR/NET — label 1). Images are pre-processed, i.e. co-registered to the same anatomical template, interpolated to the same resolution (1mm x 1mm x 1mm) and skull-stripped. Features like Age, survival days and resection status for 213 HGG images are provided separately for Overall Survival(OS). Validation dataset consists of 125 images, with the same preprocessing as well as additional features as mentioned for OS.

\section{Proposed Method}

\subsection{Task 1: Tumor Segmentation}
\begin{figure}
\includegraphics[width=\textwidth]{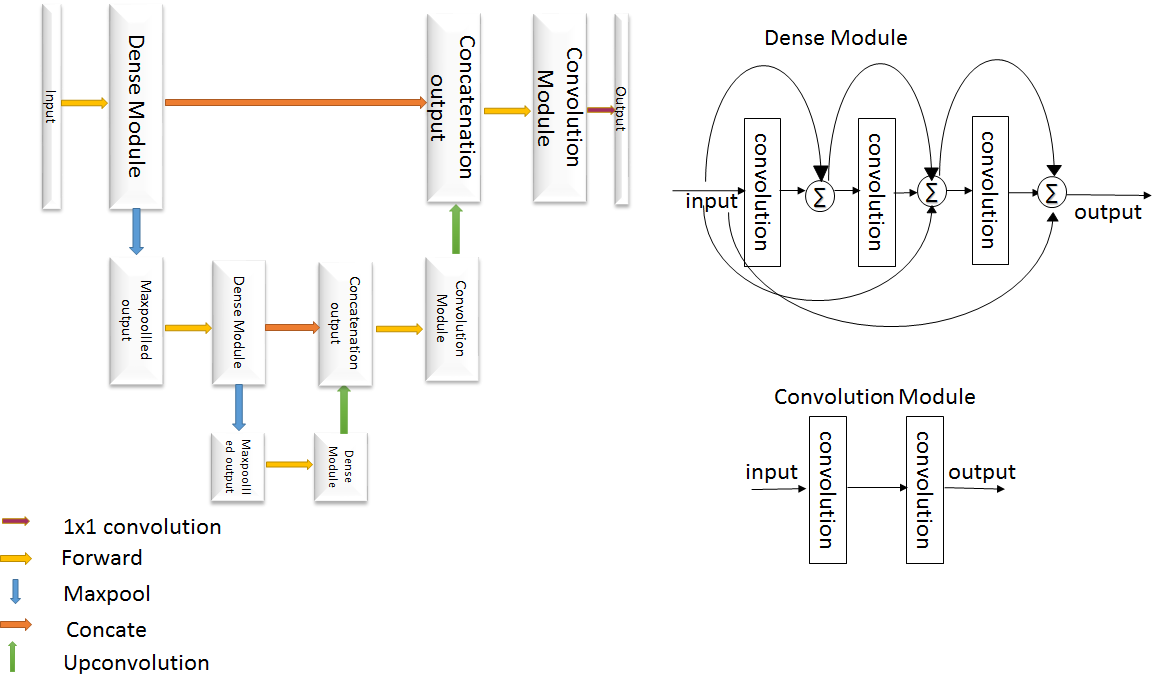}
\caption{Proposed Network Architecture} \label{fig1}
\end{figure}

Fully convolutional neural network(FCN) provides end-to-end semantic segmentation for the input of the arbitrary size and learns global information related to it. Our network is inherited from the network proposed by \cite{dong2017automatic}. The proposed network uses three-layer encoder-decoder architecture with the dense connection between the successive convolution layers and skip-connections across peer layers as shown in \ref{fig1}. The network contains three dense modules and two convolution modules. Input to the network is 2D slices of four modality(T1,T2,T1c,FLAIR) images of size 240x240. Each dense module generates 64, 128 and 256 feature maps respectively. Each convolution module generates 128 and 64 feature maps applying 1x1 convolution at the end to generate single probability map of subcomponent for which it is trained. Preprocessing includes z-score normalization of the training images. To prepare the training image set, from all the image volumes last ten slices are removed as it does not provide any information. 

Brain tumor segmentation has highly imbalanced data where tumorous slices are very less compared to non-tumorous slices, such imbalance dataset reduces network accuracy. The approach of transfer learning mentioned in\cite{pan2009survey} deals with such issue.  Authors have shown the usefulness of the transfer learning for training a network with/without labels for similar or different tasks. Initially we have trained the network for whole tumor. No. of slices are more for whole tumor compared to subregions. This step provides tumor localization in the brain. This whole tumor parameters are supplied to subregion(i.e. edema, enhancing tumor and necrotic-core) training for faster convergence and better localization.  

We have trained the network separately with two type of loss functions: soft diceloss function and focal loss.
\begin{itemize}
\item Soft Dice Loss: : is a measure to find overlap between two regions. \begin{equation}
Soft Dice Loss = 1 - \frac{2\sum_{voxels}{y_{true}}{y_{pred}}}{\sum_{voxels}{y_{pred}}^{2}+\sum_{voxels}{y_{true}}^{2}}
\end{equation}
${y_{true}}$ represents ground truth and ${y_{pred}}$ represents network output probability. Dice loss function directly considers the predicted probabilities without converting into binary output. Numerator provides common true prediction between input and target whereas denominator provides individual separate true predictions. This ratio normalizes the loss according to the target mask, and allows learning even from very small spatial representation of target mask.
\item Focal Loss\cite{lin2017focal}: balances between negative and positive samples by tuning $\alpha$ and focusing parameter $\gamma$ deals with easy and hard examples.

\begin{equation}
FL({p_{t}}) = -{\alpha_{t}}(1 - {p_{t}})^{\gamma}log({p_{t}})
\end{equation}

The modulating factor $(1 - {p_{t}})^{\gamma}$ adjusts the rate at which easy examples are down-weighted.
\end{itemize}

In both the cases, the network is trained on 75\% dataset as training images 25\% dataset as validation(part of training dataset). Evaluation results are generated based on Whole Tumor(WT), Tumor Core(TC) and Enhancing Tumor(ET). Table \ref{tab1} and Table \ref{tab2} shows various evaluation metrics of dice loss function and Table \ref{tab3} and Table \ref{tab4} are for focal loss. Challenge validation set is provided separately in addition to the training dataset. Table\ref{tab6} shows comparison of training dataset results of proposed method with average of top ten methods according to leader board. This comparison is done irrespective of multiple submission as well as without the knowledge of segmentation method used.

\begin{table}
\centering
\caption{DSC, Sensitivity and Hausdrorff95 for BraTS 2019 training dataset with dice loss.}\label{tab1}
\begin{tabular}{|c|c|c|c|c|c|c|c|c|c|}
\hline
 \multirow{2}{*}{}& \multicolumn{3}{c}{DISC} & \multicolumn{3}{c}{Sensitivity}  & \multicolumn{3}{c}{Hausdorff95} \\
 \hline
 & ET & WT & TC & ET & WT & TC  & ET & WT & TC \\
\hline
Mean & 0.74 & 0.89 & 0.85 & 0.73 & 0.83 & 0.80 & 5.42 & 6.41 & 5.82 \\
StdDev & 0.25 & 0.10 & 0.17 & 0.22 & 0.13 & 0.19 & 13.13 & 6.25 & 7.73 \\
Median & 0.83 & 0.92 & 0.90 & 0.78 & 0.87 & 0.86 & 2 & 4.90 & 4  \\
25quantile & 0.72 & 0.88 & 0.85 & 0.66 & 0.81 & 0.77 & 1.41 & 3.46 & 2.83 \\
75quantile & 0.89 & 0.94 & 0.93 & 0.87 & 0.91 & 0.90 & 3.16 & 7.31 & 6 \\
\hline
\end{tabular}
\end{table}

\begin{table}
\centering
\caption{DSC, Sensitivity and Hausdrorff95 for BraTS 2019 validation dataset with dice loss.}\label{tab2}
\begin{tabular}{|c|c|c|c|c|c|c|c|c|c|}
\hline
 \multirow{2}{*}{}& \multicolumn{3}{c}{DISC} & \multicolumn{3}{c}{Sensitivity} & \multicolumn{3}{c}{Hausdorff95} \\
 \hline
 & ET & WT & TC & ET & WT & TC & ET & WT & TC \\
\hline
Mean & 0.60 & 0.70 & 0.63 & 0.59 & 0.63 & 0.61 & 11.69 & 14.33 & 17.10 \\
StdDev & 0.33 & 0.23 & 0.30 & 0.31 & 0.25 & 0.30 & 20.31 & 18.24 & 22.33 \\
Median & 0.75 & 0.80 & 0.75 & 0.70 & 0.73 & 0.73& 3.61 & 7.81 & 8.25 \\
25quantile & 0.33 & 0.51 & 0.45 & 0.33 & 0.44 & 0.38 & 2 & 5.20 & 4.58 \\
75quantile & 0.85 & 0.88 & 0.88 & 0.84 & 0.83 & 0.87 & 10.18 & 13.45 & 16.28\\
\hline
\end{tabular};
\end{table}

\begin{table}
\centering
\caption{DSC, Sensitivity and Hausdrorff95 for BraTS 2019 training dataset with focal loss.}\label{tab3}
\begin{tabular}{|c|c|c|c|c|c|c|c|c|c|}
\hline
 \multirow{2}{*}{}& \multicolumn{3}{c}{DISC} & \multicolumn{3}{c}{Sensitivity}  & \multicolumn{3}{c}{Hausdorff95} \\
 \hline
 & ET & WT & TC & ET & WT & TC  & ET & WT & TC \\
\hline
Mean & 0.79 & 0.92 & 0.90 & 0.79 & 0.90 & 0.88 & 4.07 & 4.23 & 3.75 \\
StdDev & 0.25 & 0.09 & 0.12 & 0.21 & 0.12 & 0.14 & 11.66 & 6.39 & 7.79 \\
Median & 0.87 & 0.95 & 0.93 & 0.85 & 0.94 & 0.92 &	1.41 & 2.24 & 2 \\
25quantile & 0.81 & 0.91 & 0.89 & 0.77 & 0.89 & 0.88 & 1 & 1.41 & 1.41 \\
75quantile & 0.92 & 0.96 & 0.96 & 0.91 & 0.96 & 0.95 & 1.73 & 4.24 & 3 \\
\hline
\end{tabular}
\end{table}

\begin{table}
\centering
\caption{DSC, Sensitivity and Hausdrorff95 for BraTS 2019 validation dataset with focal loss.}\label{tab4}
\begin{tabular}{|c|c|c|c|c|c|c|c|c|c|}
\hline
 \multirow{2}{*}{}& \multicolumn{3}{c}{DISC} & \multicolumn{3}{c}{Sensitivity} & \multicolumn{3}{c}{Hausdorff95} \\
 \hline
 & ET & WT & TC & ET & WT & TC & ET & WT & TC \\
\hline
Mean & 0.59 & 0.73 & 0.65 & 0.59 & 0.67 & 0.64 & 	9.62 & 12.80 & 15.37 \\
StdDev & 0.34 & 0.24 & 0.30 & 0.33 & 0.25 & 0.31 &	15.83 & 16.86 & 19.90 \\
Median & 0.76 & 0.84 & 0.78 & 0.71 & 0.75 & 0.76 &	3.60 & 7.48 & 7.81 \\
25quantile & 0.29 & 0.65 & 0.51 & 0.33 & 0.54 & 0.41	& 1.93 & 4.58 & 4 \\
75quantile & 0.85 & 0.89 & 0.88 & 0.86 & 0.88 & 0.88 &	7.98 & 12.80 & 16.15 \\
\hline
\end{tabular}
\end{table}

\begin{table}
\centering
\caption{Comparison of DSC, Sensitivity and Hausdrorff95 for BraTS 2019 training dataset with average of top 10 teams.}\label{tab6}
\begin{tabular}{|c|c|c|c|c|c|c|c|c|c|}
\hline
 \multirow{2}{*}{ }& \multicolumn{3}{c}{DISC} & \multicolumn{3}{c}{Sensitivity} & \multicolumn{3}{c}{Hausdorff95} \\
 \hline
 & ET & WT & TC & ET & WT & TC & ET & WT & TC \\
\hline
\textbf{Average of top 10 teams} & 0.80 & 0.91 & 0.87 & 0.83 & 0.91 & 0.88 & 	3.96 & 7.54 & 7.21 \\
\hline
\textbf{Proposed} & 0.79 & 0.92 & 0.90 & 0.79 & 0.90 & 0.88 & 4.07 & 4.23 & 3.75 \\

\hline
\end{tabular}
\end{table}

\begin{figure}[!h]
\centering
\begin{subfigure}{.3\textwidth}
\includegraphics[width=4cm,height=4cm]{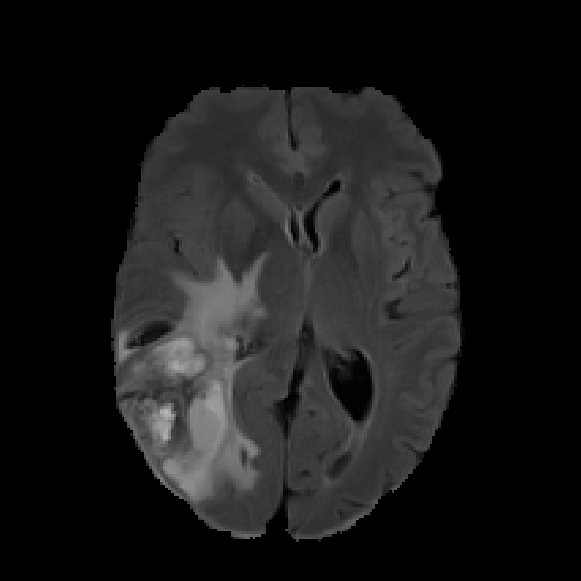}
\caption{FLAIR}
\end{subfigure}
\centering
\begin{subfigure}{.3\textwidth}
\includegraphics[width=4cm,height=4cm]{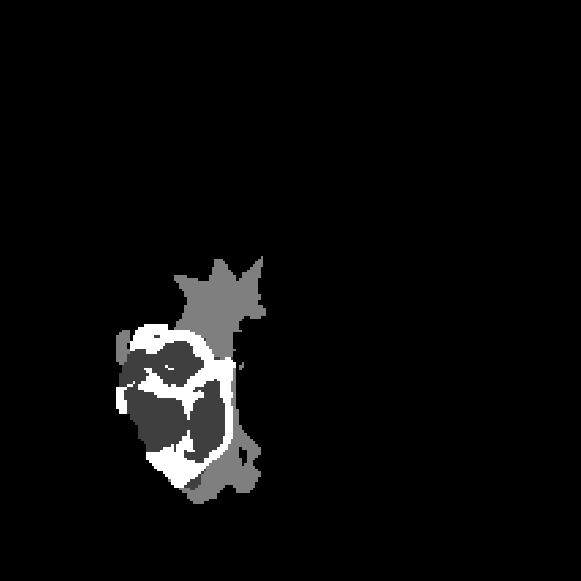}
\caption{Ground Truth}
\end{subfigure}
\centering
\begin{subfigure}{.3\textwidth}
\includegraphics[width=4cm,height=4cm]{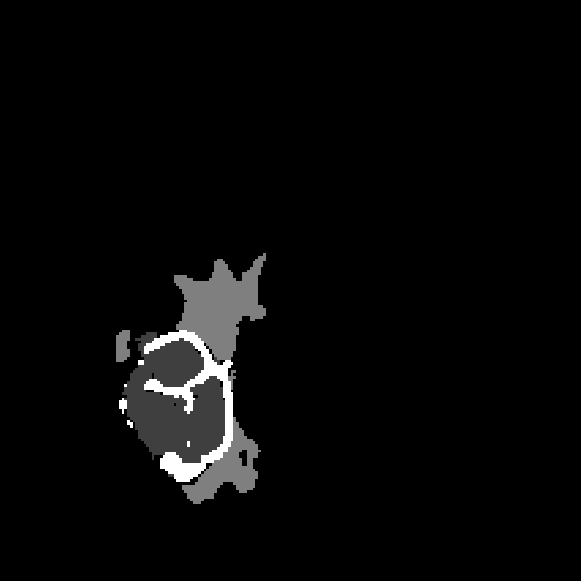}
\caption{Segmentation result}
\end{subfigure}
\caption{Segmentation result with dice loss}
\label{fig2}
\end{figure}

\begin{figure}[!h]
\centering
\begin{subfigure}{.3\textwidth}
\includegraphics[width=4cm,height=4cm]{trainingdiceloss.png}
\caption{FLAIR}
\end{subfigure}
\centering
\begin{subfigure}{.3\textwidth}
\includegraphics[width=4cm,height=4cm]{GTdiceloss.png}
\caption{Ground Truth}
\end{subfigure}
\centering
\begin{subfigure}{.3\textwidth}
\includegraphics[width=4cm,height=4cm]{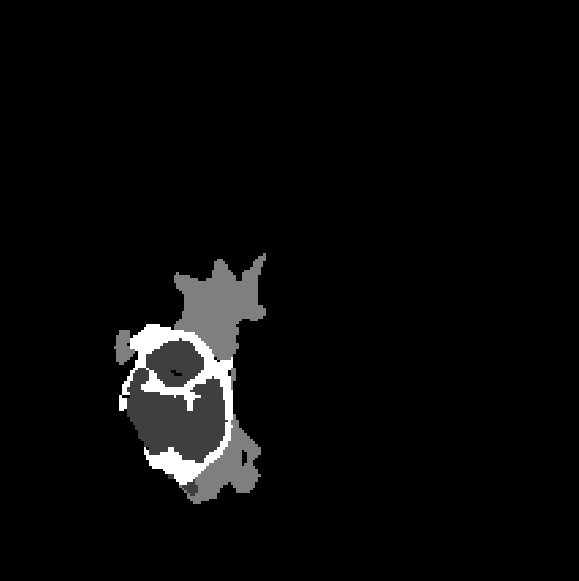}
\caption{Segmentation result}
\end{subfigure}
\caption{Segmentation result with focal loss}
\label{fig3}
\end{figure}

Fig. \ref{fig2} shows the segmentation of a tumorous slice with dice loss and Fig. \ref{fig3} shows for focal loss. From the figure, we can observe that enhancing tumor segmentation is improving in focal loss implementation compared to dice loss implementation.

\subsection{Task 2: Overall Survival prediction}
OS prediction deals with predicting no. of days for which patient will survive after providing appropriate treatment. We have used following features to train Random Forest Regressor(RFR):
\begin{itemize}
\item \textbf{Statistical Features}: amount of edema, amount of necrosis, amount of enhancing tumor, extent of tumor and proportion of tumor
\item \textbf{Radiomic Features\cite{van2017computational} for tumor core}: Elongation, flatness, minor axis length, major axis length, 2D diameter row, 2D diameter column, spherity, surface area, 2D diameter slice, 3D diameter, etc
\item \textbf{Age}(available with BraTS dataset)
\end{itemize}

Tumor core plays major role in treatment of tumors and all the provided scans are pre-operative. We have extracted shape features of tumor core using radiomics package\cite{van2017computational}, whole tumor statistical features and age which is provided with the dataset. RFR is trained on these features extracted from the 213 ground truth images. In the trained RFR, features of network segmented images are supplied and prediction is done for OS days. OS accuracy for training as well as validation dataset of the images whose resection status is Gross Total Resection(GTR) is shown in Table \ref{tab5}. 

\begin{table}
\centering
\caption{OS accuracy for training as well as validation dataset.}\label{tab5}
\begin{tabular}{|c|c|c|c|c|c|}
\hline
\textbf{Dataset} & \textbf{Accuracy} & \textbf{MSE} & \textbf{MedianSE} & \textbf{StdSE} &  \textbf{SpearmanR} \\ 
\hline
\textbf{Training} & 0.554 & 57633.216 & 9467.29 & 126525.112 & 0.657 \\
\hline
\textbf{Validation} & 0.517 & 121803.886 & 46096.09 & 161666.751 & 0.128 \\
\hline
\end{tabular}
\end{table}

According to the study \cite{sun2015integrative}, gender plays an important role in response to the tumor treatment. In addition to the feature 'Age', if 'gender' is also included in the feature list then OS accuracy can greatly improve.

\section{Conclusion}
The paper uses three-layer deep encoder-decoder architecture for semantic segmentation where at encoding side dense module and at decoding side convolution module are incorporated. The network achieves comparable DSC for training dataset with other methods of leader board but generates little poor results for validation dataset. One probable reason for it can be over-fitting of the network on the training dataset. Change in the module design or some kind of regularization may deal with over-fitting. For OS prediction age, statistical, and necrosis shape features are considered to train RFR. RFR achieves comparable accuracy with other leader board methods.

\section*{Acknowledgement}

The authors would like to thank NVIDIA Corporation for donating the Quadro K5200 and Quadro P5000 GPU used for this research, Dr. Krutarth Agravat (Medical Officer, Essar Ltd) for clearing our doubts related to medical concepts. The authors acknowledge continuous support from Professor Sanjay Chaudhary, Professor Manjunath Joshi and Professor N. Padmanabhan for this work.

%
%
%

\bibliographystyle{splncs04}
\bibliography{bibtex_example}

\begin{thebibliography}{10}
\providecommand{\url}[1]{\texttt{#1}}
\providecommand{\urlprefix}{URL }
\providecommand{\doi}[1]{https://doi.org/#1}

\bibitem{agravat2019os}
Agravat, R.R., Raval, M.S.: Prediction of overall survival of brain tumor
  patients, submitted

\bibitem{agravat2018deep}
Agravat, R.R., Raval, M.S.: Deep learning for automated brain tumor
  segmentation in mri images. In: Soft Computing Based Medical Image Analysis,
  pp. 183--201. Elsevier (2018)

\bibitem{akbari2014pattern}
Akbari, H., Macyszyn, L., Da, X., Wolf, R.L., Bilello, M., Verma, R.,
  O’Rourke, D.M., Davatzikos, C.: Pattern analysis of dynamic susceptibility
  contrast-enhanced mr imaging demonstrates peritumoral tissue heterogeneity.
  Radiology  \textbf{273}(2),  502--510 (2014)

\bibitem{badrinarayanan2017segnet}
Badrinarayanan, V., Kendall, A., Cipolla, R.: Segnet: A deep convolutional
  encoder-decoder architecture for image segmentation. IEEE transactions on
  pattern analysis and machine intelligence  \textbf{39}(12),  2481--2495
  (2017)

\bibitem{baid2018deep}
Baid, U., Talbar, S., Rane, S., Gupta, S., Thakur, M.H., Moiyadi, A., Thakur,
  S., Mahajan, A.: Deep learning radiomics algorithm for gliomas (drag) model:
  A novel approach using 3d unet based deep convolutional neural network for
  predicting survival in gliomas. In: International MICCAI Brainlesion
  Workshop. pp. 369--379. Springer (2018)

\bibitem{bakas2017segmentation}
Bakas, S., Akbari, H., Sotiras, A., Bilello, M., Rozycki, M., Kirby, J.,
  Freymann, J., Farahani, K., Davatzikos, C.: Segmentation labels and radiomic
  features for the pre-operative scans of the tcga-gbm collection. the cancer
  imaging archive (2017) (2017)

\bibitem{bakas2017segmentation1}
Bakas, S., Akbari, H., Sotiras, A., Bilello, M., Rozycki, M., Kirby, J.,
  Freymann, J., Farahani, K., Davatzikos, C.: Segmentation labels and radiomic
  features for the pre-operative scans of the tcga-lgg collection. The Cancer
  Imaging Archive  \textbf{286} (2017)

\bibitem{bakas2017advancing}
Bakas, S., Akbari, H., Sotiras, A., Bilello, M., Rozycki, M., Kirby, J.S.,
  Freymann, J.B., Farahani, K., Davatzikos, C.: Advancing the cancer genome
  atlas glioma mri collections with expert segmentation labels and radiomic
  features. Scientific data  \textbf{4},  170117 (2017)

\bibitem{bakas2018identifying}
Bakas, S., Reyes, M., Jakab, A., Bauer, S., Rempfler, M., Crimi, A., Shinohara,
  R.T., Berger, C., Ha, S.M., Rozycki, M., et~al.: Identifying the best machine
  learning algorithms for brain tumor segmentation, progression assessment, and
  overall survival prediction in the brats challenge. arXiv preprint
  arXiv:1811.02629  (2018)

\bibitem{chen2018drinet}
Chen, L., Bentley, P., Mori, K., Misawa, K., Fujiwara, M., Rueckert, D.: Drinet
  for medical image segmentation. IEEE transactions on medical imaging
  \textbf{37}(11),  2453--2462 (2018)

\bibitem{dong2017automatic}
Dong, H., Yang, G., Liu, F., Mo, Y., Guo, Y.: Automatic brain tumor detection
  and segmentation using u-net based fully convolutional networks. In: annual
  conference on medical image understanding and analysis. pp. 506--517.
  Springer (2017)

\bibitem{he2016deep}
He, K., Zhang, X., Ren, S., Sun, J.: Deep residual learning for image
  recognition. In: Proceedings of the IEEE conference on computer vision and
  pattern recognition. pp. 770--778 (2016)

\bibitem{iandola2014densenet}
Iandola, F., Moskewicz, M., Karayev, S., Girshick, R., Darrell, T., Keutzer,
  K.: Densenet: Implementing efficient convnet descriptor pyramids. arXiv
  preprint arXiv:1404.1869  (2014)

\bibitem{kao2018brain}
Kao, P.Y., Ngo, T., Zhang, A., Chen, J.W., Manjunath, B.: Brain tumor
  segmentation and tractographic feature extraction from structural mr images
  for overall survival prediction. In: International MICCAI Brainlesion
  Workshop. pp. 128--141. Springer (2018)

\bibitem{lin2017focal}
Lin, T.Y., Goyal, P., Girshick, R., He, K., Doll{\'a}r, P.: Focal loss for
  dense object detection. In: Proceedings of the IEEE international conference
  on computer vision. pp. 2980--2988 (2017)

\bibitem{menze2014multimodal}
Menze, B.H., Jakab, A., Bauer, S., Kalpathy-Cramer, J., Farahani, K., Kirby,
  J., Burren, Y., Porz, N., Slotboom, J., Wiest, R., et~al.: The multimodal
  brain tumor image segmentation benchmark (brats). IEEE transactions on
  medical imaging  \textbf{34}(10),  1993--2024 (2014)

\bibitem{milletari2016v}
Milletari, F., Navab, N., Ahmadi, S.A.: V-net: Fully convolutional neural
  networks for volumetric medical image segmentation. In: 2016 Fourth
  International Conference on 3D Vision (3DV). pp. 565--571. IEEE (2016)

\bibitem{pan2009survey}
Pan, S.J., Yang, Q.: A survey on transfer learning. IEEE Transactions on
  knowledge and data engineering  \textbf{22}(10),  1345--1359 (2009)

\bibitem{ronneberger2015u}
Ronneberger, O., Fischer, P., Brox, T.: U-net: Convolutional networks for
  biomedical image segmentation. In: International Conference on Medical image
  computing and computer-assisted intervention. pp. 234--241. Springer (2015)

\bibitem{sun2015integrative}
Sun, T., Plutynski, A., Ward, S., Rubin, J.B.: An integrative view on sex
  differences in brain tumors. Cellular and molecular life sciences
  \textbf{72}(17),  3323--3342 (2015)

\bibitem{van2017computational}
Van~Griethuysen, J.J., Fedorov, A., Parmar, C., Hosny, A., Aucoin, N., Narayan,
  V., Beets-Tan, R.G., Fillion-Robin, J.C., Pieper, S., Aerts, H.J.:
  Computational radiomics system to decode the radiographic phenotype. Cancer
  research  \textbf{77}(21),  e104--e107 (2017)

\end{thebibliography}
\end{document}